\documentclass[paper]{jpsj3}
\usepackage{txfonts}
\usepackage{color}
\newcommand{\figwidth}{3.375in}

\title{Statistical Mechanical Models of Integer Factorization Problem}

\author{Chihiro H. Nakajima\thanks{E-mail address: chihiro.nakajima.d3@tohoku.ac.jp} and Masayuki Ohzeki\thanks{E-mail address: mohzeki@i.kyoto-u.ac.jp}}

\inst{$^{1}$\address{WPI-AIMR, Tohoku University, 2-1-1 Katahira, Aoba-ku, Sendai 980-8577, Japan}
$^{2}$\address{Department of Systems Science, Graduate School of Informatics, Kyoto University, 36-1 Yoshida Hon-machi, Sakyo-ku, Kyoto 606-8501, Japan}
}
\abst{We formulate the integer factorization problem via a formulation of the searching problem for the ground state of a statistical mechanical Hamiltonian.
The first passage time required to find a correct divisor of a composite number signifies the exponential computational hardness.
The analysis of the density of states of two macroscopic quantities, i.e., the energy and the Hamming distance from the correct solutions, leads to the conclusion that the ground state (correct solution) is completely isolated from the other low-energy states, with the distance being proportional to the system size.
In addition, the profile of the microcanonical entropy of the model has two peculiar features that are each related to two marked changes in the energy region sampled via Monte Carlo simulation or simulated annealing.
Hence, we find a peculiar first-order phase transition in our model.}

\kword{Statistical mechanics, Monte Carlo simulation, Prime factorization, Quantum annealing}

\begin{document}
\maketitle

\section{Introduction}
The relationship between the hardness (i.e., complexity) of computational problems and the behavior of their corresponding statistical mechanical models has attracted considerable research attention in recent decades.
The investigation of problems such as the 3-satisfiability problem (3SAT) \cite{MZKST}, number partitioning \cite{Mer}, vertex covering \cite{WH}, and graph coloring \cite{KMRSZ} has highlighted the relationship between the spin-glass transition, or the structure of the energy landscape, and the behavior of the average-case computational complexity.
By using these approaches, previous studies have examined nondeterministic polynomial time (NP) complete or equivalently hard NP-hard problems.
However, attempts to apply the spin-glass-based theory to problems outside of the class of NP-complete scenarios have attracted less attention.
The integer factorization problem is one of the computational problems categorized as being of a certain class separate from NP-complete problems, and this problem is considered to be of a different hardness to NP-complete problems.
In fact, in the field of computational complexity theory, the factorization problem is categorized as being of the $\mathrm{NP} \cap \mathrm{co}\rule[1mm]{1.mm}{0.3mm}\mathrm{NP}$ class \cite{AKS}.
If this problem were NP-complete, however $\mathrm{NP}=\mathrm{co}\rule[1mm]{1.mm}{0.3mm}\mathrm{NP}$ would be concluded.
The $\mathrm{NP}  \ \mathrm{vs}  \ \mathrm{co}\rule[1mm]{1.mm}{0.3mm}\mathrm{NP}$ problem is considered to be as formidable as the $\mathrm{P}  \ \mathrm{vs}  \ \mathrm{NP}$ problem.
In fact, the factorization problem can be solved in quasi-exponential time on average using certain algorithms, such as the elliptic curve method \cite{Lst} or the number sieve method \cite{LL}.
Furthermore, the spin-glass approach to the integer factorization problem provides a penetrative insight into further consequences of average complexity.

The integer factorization problem itself is also of interest in the field of quantum computation.
Shor has proposed a quantum algorithm that can solve this problem in polynomial time\cite{Shor}.
However, controversy exists as to which aspect of quantum mechanics is crucially responsible for the obtained acceleration.
Investigating the behaviors of various quantum algorithms while focusing on a particular problem leads to broader viewpoints from which quantum properties in computing can be reconsidered.
For example, quantum annealing \cite{KN,MN,ON} is an alternative algorithm, which utilizes fictitious quantum fluctuation to find an optimal solution by searching for the minimum of the Hamiltonian in question.
The quasi-adiabatic behavior in the quantum annealing can be mapped into a quasi-equilibrium property with the common Hamiltonian \cite{TYN}.
As regards the estimation of quantum annealing efficiencies, the classical and quantum phase-transition properties of corresponding statistical mechanical models provide useful insights.
On the basis of the size dependence of the minimum energy gap, which characterizes the quantum phase transition, and the various consequences of the adiabatic theorem, extensive studies have been conducted on the relationship between computational hardness and phase transitions \cite{YNS,JKK}.
As shown in a series of previous studies, the analysis of a statistical mechanical model in equilibrium can be a witness of the computational power of its quantum annealing counterpart\cite{SN,TYN}.

The statistical mechanical study of the factorization problem involves a crossover of two research avenues: spin-glass theory and quantum annealing.
In this present study, we formulate the integer factorization problem as a combinatorial optimization problem and treat it as a statistical mechanical model.
The structure of the energy landscape and the phase transitions, as monitored from the density of states and the specific heat, are discussed in detail.

The organization of the present paper is as follows:
In the Sect.2, the integer factorization problem is formulated as a statistical mechanical model.
In the Sect.3, we perform numerical experiments to investigate the computational complexity of the integer factorization problem by Monte-Carlo simulation.
Hence, we identify a peculiar phase-transition behavior in the statical mechanical model.
In the Sect.4, we analyze the phase transition in terms of various factors.
The section 5 is devoted to a discussion of the future direction of our present topics of research.

\section{Models and Implementations}
Let $N = p_1 \times \cdots p_l \times \cdots p_m$ be a composite number that can be factorized into prime numbers $p_1, \cdots ,p_m$.
We formulate the problem of finding a divisor (in general, regardless of whether or not the divisor is a prime number) of $N$ in terms of statistical mechanical models.
A situation in which $N$ is divided by a randomly chosen integer $d$ is considered.
First, $d$ is represented by a combination of binary variables $s_1,\cdots,s_n \in \{0,1\}$, such that
\begin{eqnarray}
d\big(\{s_i\}\big)=2+\sum_{i=0}^{n-1}s_i2^i.
\end{eqnarray}
Then, $d$ can take any integer value in $\{2,3,\cdots,2^n+1\}$ through the combination of $\{s_i\}$.
We can regard $n=\lceil  \ \log_2 \sqrt{N}  \ \rceil$ as the system size of the model, where $\lceil a \rceil$ means the smallest integer greater than the real number $a$.
Second, we impose extensivity with $n$ on the cost function (Hamiltonian) $H$; this allows the thermodynamic quantities or phase-transition phenomena to be discussed naturally.
Indeed, under this condition, the internal energy is extensive, i.e., $\langle H \rangle_{\beta} \propto n$, where the brackets denote the thermal average.
As an additional condition, the Hamiltonian takes its lowest value, $H(\{s_i\})=0$, if and only if $d$ is one of the correct divisors of $N$; otherwise, $H(\{s_i\})>0$.
We consider Hamiltonians that satisfy the above requirements by focusing on the remainder of $N$ divided by $d$,
\begin{eqnarray}\label{eq:spin_variable}
\mathrm{mod}(N,d)=\sum_{j=0}^{n-1}\sigma_j2^j,
\end{eqnarray}
where $\sigma_j \in \{0,1\}$, and by considering the maximum digits or the sum of the binary expansion coefficient.
We define the maximum-digit-based model
\begin{eqnarray}\label{eq:max_model}
H(\{s_i\})=\lceil\log_2\big(1+\mathrm{mod}(N,d)\big)\rceil,
\end{eqnarray}
and the summation-based model
\begin{eqnarray}\label{eq:sum_model}
H(\{s_i\})=\sum_{j=0}^{n-1}\sigma_j.
\end{eqnarray}
The microscopic state $d$ is represented by binary (or spin) variables $\{s_i\}$.

In this paper, we consider the cases in which $N$ is composed of two large prime numbers, $p_1$ and $p_2$ (let $p_1 < p_2$).
$p_1$ and $p_2$ are chosen such that $\lceil\log_2 p_1\rceil=n+1$ and $\lceil\log_2 p_2\rceil=n-1$.
In these cases, the model has a unique ground state.

Supposing that we find $d$ corresponding to any correct divisor of a given $N$ by updating each variable $s_i$, we then measure the density of states over two macroscopic quantities, the energy $H\big(\{s_i\}\big)$ and the Hamming distance, which is defined as
\begin{eqnarray}
\hat{Q}\big(\{s_i\}\big)=\sum_{i=0}^{n-1}\frac{1-(2s_i-1)(2s_i^{*}-1)}{2}.
\end{eqnarray}
Here, $\{s_i^{*}\}$ is the binary representation of the correct divisor.
The quantity $\hat{Q}\big(\{s_i\}\big)$ is analogous to the overlap function in the context of the spin-glass theory.
Each elemental process of local updating is changing the Hamming distance by 1.
In other words, $\hat{Q}\big(\{s_i\}\big)$ simply indicates the smallest number of flips required to reach a ground state.

We measure the density of states that take the values $\hat{Q}\big(\{s_i\}\big)=Q$ and $\hat{H}\big(\{s_i\}\big)=E$,
\begin{eqnarray}
W(E,Q)&=&\sum_{\{s_i\}}\delta\big(\hat{Q}\big(\{s_i\}\big)-Q\big)\delta\big(H\big(\{s_i\}\big)-E\big) \\
&=&\exp(S(E,Q)), \nonumber
\end{eqnarray}
where $S(E,Q)$ is the microcanonical entropy as a function of $(E,Q)$.
Using $W(E,Q)$, 
we define the probability of finding $(E,Q)$,
\begin{eqnarray}\label{eq:ensemble_r}
P(E,Q | \beta)=\frac{\sum_{E'=0}^{n}\sum_{Q'=0}^{n} \ \delta\big(Q'-Q\big)\delta\big(E'-E\big) \ W(E',Q')\exp(-\beta E')}{\sum_{E'=0}^{n}\sum_{Q'=0}^{n} W(E',Q')\exp(-\beta E')},
\end{eqnarray}
and the statistical ensemble, $\langle \cdots \rangle_{\beta}=\sum_{E=0}^{n}\sum_{Q=0}^{n}(\cdots)P(E,Q | \beta)$ with fixed temperature.
In addition, we discuss in this paper the sample average, averaged over various instances of $N$, of the quantities.
The average over $N$ is noted as $[\cdots]$.

For the numerical simulation, we employ the replica-exchange Monte Carlo method\cite{HN} in order to sample the histogram on $(E,Q)$ at each temperature $T=1/\beta$ and $W(E,Q)$.
The simulation method is explained in detail in the Appendix.

\section{Numerical Results}\label{sec:Numerical_Results}
\begin{figure}[h]
\begin{center}
\includegraphics[width=\figwidth]{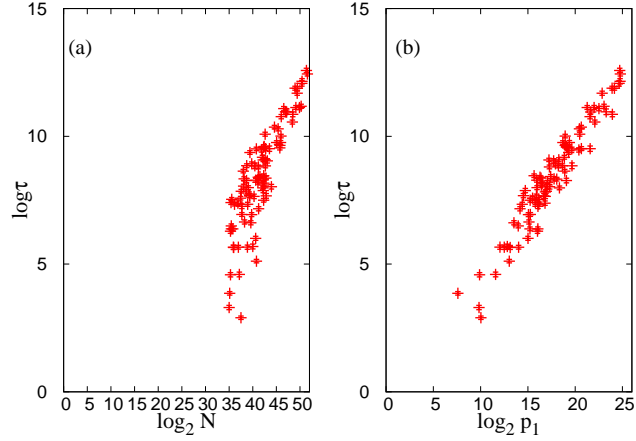}
\caption{\label{fig:epsart}First passage time $\tau$ for maximum-digit model with various instances of $N$.
(a) $\log\tau$ with respect to $\log_2 N=\log_2 p_1p_2$. (b) $\log\tau$ with respect to the logarithm of the correct divisor, $\log_2 p_1$. In both figures, the vertical axis represents the logarithm of $\tau$. Each point is the average value for more than $360$ simulations.
}
\end{center}
\end{figure}
In this and the next section, basically, the result for the Hamiltonian (\ref{eq:max_model}) are reported. Those for the Hamiltonian (\ref{eq:sum_model}) is reported in the last part of Sect. \ref{sec:entropy_profile} for comparison.
The $n$ dependence of the first passage time $\tau$ is shown in Fig. \ref{fig:epsart}, where an exponential dependence on $n$ is apparent.
This result is consistent with preliminary investigations \cite{CHN}.
To understand the computational difficulties as indicated by $\tau$ from the perspective of statistical mechanics, we further perform the numerical sampling of the static quantities up to $N \simeq 2^{512}$, i.e., for $n = 256$.

\begin{figure}[t]
\begin{center}
\includegraphics[width=\figwidth]{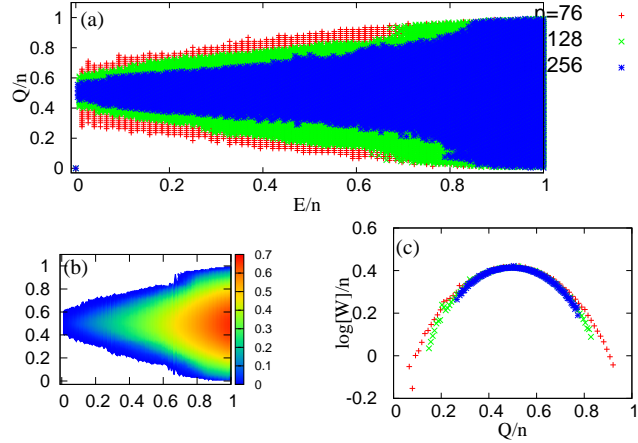}
\caption{\label{fig:xygap_max}Distribution of $\log[W]/n$ on $(E/n,Q/n)$-plane for maximum-digit model.
(a) Overplot of cases with $n=76, 128, \mathrm{and} \ 256$.
In the blank region, $W(E,Q)=0$ for all instances included in the average. 
(b) $\log[ \ W(E,Q) \ ]/n$ for $n=128$. The value is represented using a color scale.
(c) Cross section of $\log[ \ W(E,Q) \ ]/n$ at $E/n=0.6$ with $n=76, 128, \mathrm{and} \ 256$.
}
\end{center}
\end{figure}
In Fig.  \ref{fig:xygap_max}, the value of $\log[ \ W(E,Q) \ ]/n$, the annealed average, is plotted on the $(E/n,Q/n)$ plane.
To carefully discuss the region with $[W(E,Q)]=0$ on the plane, here, we adopt the annealed average, namely, taking the instance average before taking the logarithm for Fig. \ref{fig:xygap_max}.
The blank region means the region with $[ \ W(E,Q) \ ]=0$, where no microscopic state is found at any $N$ instances numerically sampled. 
In this figure, the states with low, but positive, energy are separated from the ground state by $Q$, of the order of $n$, as shown in Fig. \ref{fig:xygap_max}.
Furthermore, the states that are close to the ground state in terms of $Q$ have finite energy differences, which are also proportional to $n$.
This behavior is commonly seen at various system sizes with up to $n=256$ and various $N$ values.
Therefore, a barrier impedes the determination of the ground state at low temperatures, whereas, at high temperatures, this is simply difficult because of the rarity of the solution.
\begin{figure}[h]
\begin{center}
\includegraphics[width=\figwidth]{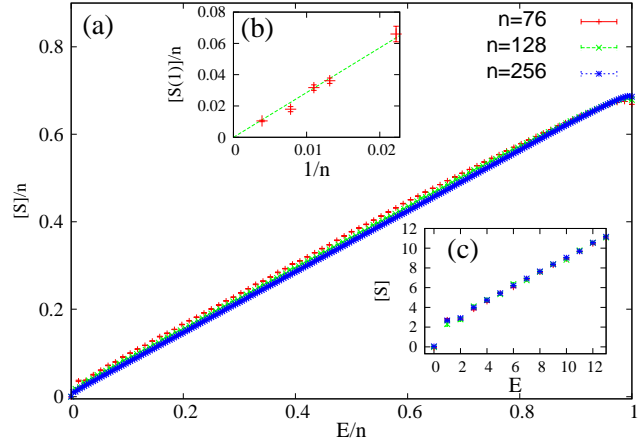}
\caption{\label{fig:ham_tmp_pfor}$S(E)$ averaged over instances, $[S(E)]=\Big[\log\Big( \sum_{Q=0}^{n}W(E,Q)\Big) \Big]$ for maximum-digit model. (a) Entire $[S]/n$ versus $E/n$ profile with $n=76, 128, \mathrm{and} \ 256$. (b) $n$ dependence of $[S(1)]/n$ versus $1/n$ for $n=46, 64, 91, 128, \mathrm{and} \ 256$. The data points are fit by a line with its gradient $\simeq 2.86 \pm 0.09$. (c) Profile of the low-$E$ region of $[S]$ versus $E$.
}
\end{center}
\end{figure}
\begin{figure}[h]
\begin{center}
\includegraphics[width=\figwidth]{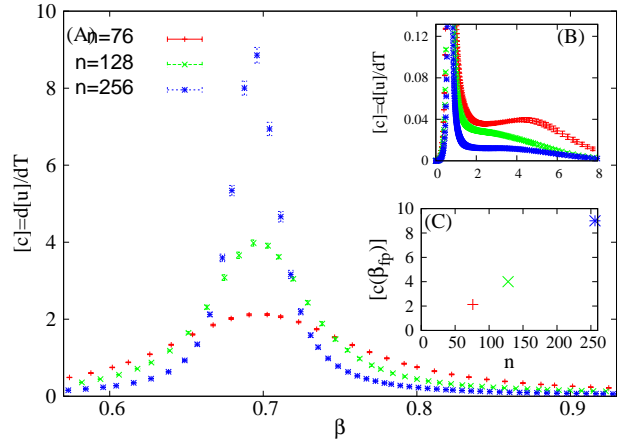}
\caption{\label{fig:sp_ht_max}$\beta$ dependence of $[c(\beta)]$ for maximum-digit model. (a) $[c(\beta)]$ profiles with respect to $\beta$ with $n=76, 128, \mathrm{and} \ 256$ around the first peak. (b) Details of $[c(\beta)]$ around the region much smaller than the first peaks. (c) First peak height with respect to $n$.
}
\end{center}
\end{figure}

The characteristics of the microcanonical entropy of this problem reveal two marked changes in the most-sampled energy region.
One may expect that the phase transitions occur at these temperatures.
Therefore, we investigate the phenomena associated with each change in the sampled region in terms of both $[S(E)]=[\log \sum_{E}W(E,Q)]$ and the specific heat $[c(\beta)]$, as shown in Figs. \ref{fig:ham_tmp_pfor} and \ref{fig:sp_ht_max}.

One of the characteristic features apparent in the $[S(E)]$ profile is a convex ``kink''-like discontinuity of the gradient at $E=1$, as shown in Fig. \ref{fig:ham_tmp_pfor}(c).
On each side of this convex kink, the $\partial [S(E)]/\partial E$ gradients are discontinuously different.
As the gradients correspond to each microcanonical temperature $1/\beta$, this discontinuity induces the phenomenon that the internal energy $[\langle E \rangle_{\beta}]$ is approximately a constant ( at $\simeq 1$) in the finite-temperature region $1/\beta_{\mathrm{kl}} < T < 1/\beta_{\mathrm{kh}}$, where $\beta_{\mathrm{kl}}$ and $\beta_{\mathrm{kh}}$ are the microcanonical temperatures corresponding to the gradients on the lower- and higher-$E$ sides of the kink, respectively.
The average values of $\beta_{\mathrm{kh}}$ and $\beta_{\mathrm{kl}}$ are roughly $\simeq \log 2$ and $\simeq 2.86$, respectively, where $\beta_{\mathrm{kh}}$ and $\beta_{\mathrm{kl}}$ are estimated from the gradient of the slope and $\big( \ [S(1)]-[S(0)] \ \big)/\big(1-0\big)$, respectively.
As the change in $[\langle E \rangle_{\beta}]$ is small, the value of $c(\beta)$ is small in this region, and the change appears as a dip in the $c$ profile.
For each $N$, the corresponding microcanonical inverse temperatures have a rather sensitive dependence.
Hence, dips in the $c$ profile for each $N$ seems to be leveled in sample average and appear as a shoulder in $[c(\beta)]$, as shown in Fig. \ref{fig:sp_ht_max}(b).

\begin{figure}[h]
\begin{center}
\includegraphics[width=\figwidth]{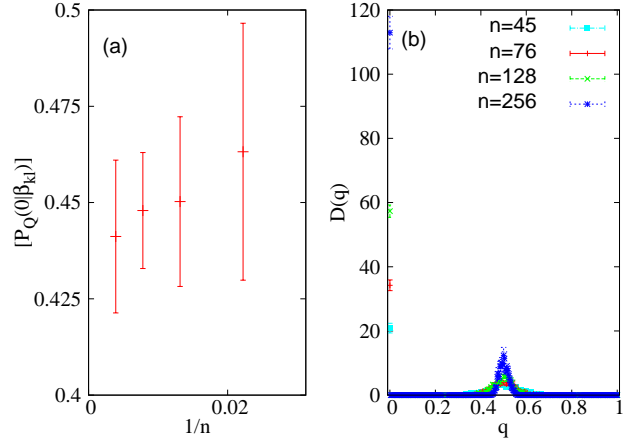} 
\caption{\label{fig:kink_low}(a) Sample-averaged probability of finding $Q=0$ at $T=1/\beta_{\mathrm{kl}}$ for each $n$, $[P_Q(0 \ | \ \beta_{\mathrm{kl}})]$. $\beta_{\mathrm{kl}}$ for each $n$ is determined from $\beta_{\mathrm{kl}}=([S(1)]-[S(0)])/(1-0)$. (b) Profile of $D(q)$ at $T=1/\beta_{\mathrm{kl}}$ for each $n$.}
\end{center}
\end{figure}

The distribution of the scaled Hamming distance, which is defined as  $D(q)=\frac{P_Q(Q \ | \ \beta)}{\Delta q}=(1/\Delta q)\sum_{E=0}^n P(E,Q \ | \ \beta)$, where the bin width $\Delta q$ is set to be $\Delta q=1/n$ and $q=Q/n$, has a two-peak structure for each $n$, as shown in Fig. \ref{fig:kink_low}(b).
In addition, the average probability of finding $Q=0$, $[ P_Q(0 \ | \ \beta) ]$, becomes approximately $0.5$ near the temperature region $T \simeq 1/\beta_{\mathrm{kl}}$, as shown in Fig. \ref{fig:kink_low}(a).
These behaviors indicate that the macroscopic states governing the statistical ensemble change with the macroscopic shifting of their $Q$ below and above $T = 1/\beta_{\mathrm{kl}}$.
This is one of the previously mentioned marked changes in the sampled states apparent in the Monte Carlo simulation.

Another feature of$[S(E)]$ is a roughly linear slope over a wide range of $E$.
In the $[S(E)]$ profile, the peculiar linear slope continues from low- to high-$E$ regions.
The value of the gradient of this slope approaches $\beta_{\mathrm{fp}} = \log(2)$ with increasing $n$.
At $T=1/\beta_{\mathrm{fp}}$, the system exhibits a jump in $[\langle E \rangle_{\beta}]$.
We confirm the emergence of a peak in the profile of $[c(\beta)]$ involved in this jump.
The average $c(\beta)$ over various $N$ values, $[c(\beta)]=(\beta^2/n)\Big[ \ \langle \big(E - \langle E \rangle_{\beta} \big)^2 \rangle_{\beta} \ \Big]$, is shown in Fig. \ref{fig:sp_ht_max}.
The peak seems to approach a delta peak with increasing $n$, as shown in Fig. \ref{fig:sp_ht_max}(a).
The behaviors of $[S(E)]$ and $[c(\beta)]$ may indicate a first-order phase transition. 
However, detailed analysis of this slope reveals a different behavior from that of an ordinary first-order transition.

\section{Linear Microcanonical Entropy Profile}\label{sec:entropy_profile}
To understand the phase-transition behavior corresponding to the slope in $[S(E)]$,
we utilize a simple model to demonstrate the behavior of the thermodynamic quantity involved in the linear profile of $S(E)$.
In this section, on the basis of the random-energy model, we compute the free energy of the simplified model of the factorization problem.
The $E$ of the random-energy model is generated in accordance with the Gaussian distribution.
However, instead of the Gaussian distribution, we define the probability distribution of $E$ as
\begin{equation}
P(E) \propto \exp\left( a E \right)\theta\left( E < b n\right),
\end{equation}
where $\theta(x) = 1~(x>0)$ and $0~(x\le 0)$, $a$ is a coefficient of the linear shape of the microcanonical entropy, and $b$ indicates the maximum value of $E/n$.
The probability distribution of $E$, $P(E)$, is related to the density of states according to
\begin{equation}
\frac{1}{2^n}\sum_{\{s_i\}}\delta\left( E - H_{\rm eff}(\{s_i\})\right) = P(E),
\end{equation}
where we define the effective Hamiltonian of our model as $H_{\rm eff}(\{s_i\})$.
In other words, the definition of the probability distribution reflects on the linear shape of the microcanonical entropy.
We evaluate the partition function following the description of the statistical mechanics as
\begin{equation}
Z = \sum_{\{s_i\}}\exp\left( - \beta H(\{s_i\})\right).
\end{equation}
We then substitute the identity of the integral of the delta function $1=\int dE\delta(E-H_{\rm eff}(\{s_i\}))$ into the definition of $Z$ and obtain
\begin{equation}
Z = \sum_{\{s_i\}} \int_0^b d(ne)\exp\left\{n(\log 2 + a e - \beta e)\right\},
\end{equation}
where $e = E/n$.
The saddle-point method leads to the free energy per single spin $f$ as
\begin{equation}
-\beta f = \left\{
\begin{array}{ll}
\log 2 +(a - \beta)b,& (\beta < a), \\
\log 2,  & (\beta \ge a).
\end{array}
\right.
\end{equation}
This computation is identical to the case of the random-energy model in the context of the spin-glass theory.
In the model (\ref{eq:max_model}), the slope value takes $a = \log 2$.
The normalized internal energy of this random-energy model, $e^{*}$, is 
\begin{equation}
e^{*} = \left\{
\begin{array}{ll}
1, & (\beta < a), \\
0, & (\beta \ge a).
\end{array}
\right.
\end{equation}
Therefore, the linear slope of $S(E)$ in the model (\ref{eq:max_model}) exhibits a discontinuous jump in $[\langle E \rangle_{\beta}]$.
Furthermore, $[c(\beta)]$ has an infinitely strong peak at the transition point.
We should state that the phase transition at $\beta_{\rm fp} = a ( \ =\log 2)$ is the first-order transition.
Indeed, we have confirmed the discontinuous jump in $[\langle E \rangle_{\beta}]$ from the numerical results, and also confirmed the peak of $[c(\beta)]$.
Note that the above calculation was performed on a simplified model with a linear slope in $S(E)$.
We just elucidate the 
feature of the thermodynamic behavior of the factorization problem.
It would be interesting to seek a spin model with the same property as the above simplified model, similar to the case of the random-energy model, which corresponds to the $p$-body interacting spin-glass model \cite{REM,REM2}.

\begin{figure}[h]
\begin{center}
\includegraphics[width=\figwidth]{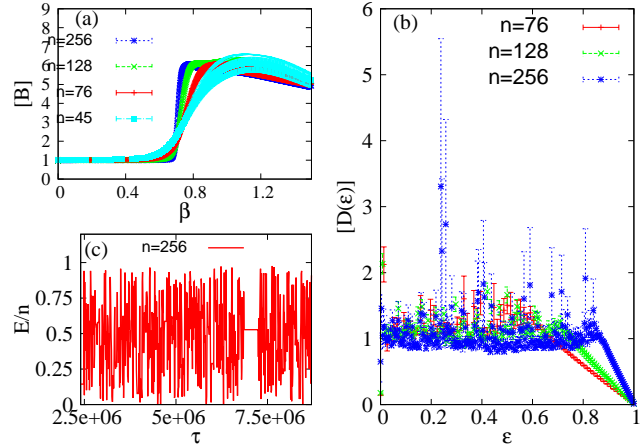}
\caption{\label{fig:can_max_fp}(a) Temperature dependence of the sample-averaged Binder ratio $[B]$ of the max model with each $n$. (b) Sample-averaged energy-density distribution $[D(\epsilon)]$ of the max model at $\beta=\beta_{\mathrm{fp}}$, which is determined by the peak of $[c(\beta)]$, with each $n$. (c) Time-step sequences of instantaneous energy per spin $E/n$ with simple Metropolis dynamics with $n=256$ for a single instance. $p_1=30640731147833253864896995273045097514997268841224331063914347087374850016503$ and $p_2=116117331593301981716052854486558696641288093228789514356535703488930327342663$}
\end{center}
\end{figure}

We focus on the phase transition of the factorization problem below, while comparing the findings to the conventional behavior of the first-order phase transition.
Let us examine the behavior of the specific heat closely, so as to determine the occurrence of the first-order phase transition.
We show the profile of the sample-averaged specific heat $[c(\beta)]$ in Fig. \ref{fig:sp_ht_max}.
The height of the first peak of $[c(\beta)]$ grows proportionally to $n$,
while the values in the vicinity of the peak decrease [see Fig. \ref{fig:sp_ht_max}(a)].
As shown in Fig. \ref{fig:can_max_fp}(a), the $\beta$ dependence of the Binder ratio $[B]= \langle E^4 \rangle/ \langle E^2 \rangle^2$ exhibits a non decaying peak at the (inverse) transition temperature. 
This behavior indicates that there is a macroscopic spread in the energy density distribution at this $\beta$.
This peak and the size dependence of the peak in $[c(\beta)]$ confirm that a macroscopic jump of the internal energy occurs with this transition.

\begin{figure}[h]
\begin{center}
\includegraphics[width=\figwidth]{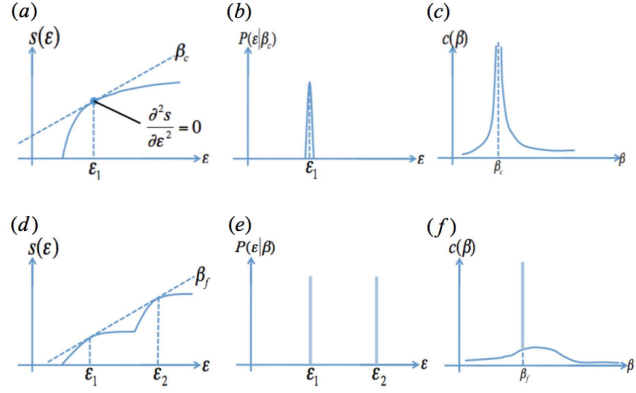} 
\caption{\label{fig:sch_fo_and_so}Schematic representation of the microcanonical entropy density [(a) and (c)], energy distribution at the transition temperature [(b) and (e)], and the specific heat $s(\epsilon), P(\epsilon | \beta_c), \mathrm{and} c(\beta)$ [(c) and (f)] with first- and second-order phase transitions. The upper part ([a), (b), and (c)] shows the cases with the second-order transition. The lower part [(d), (e), and (f)] is corresponding to the first-order transition.
}
\end{center}
\end{figure}

However, the linear slope of $[S(E)]$ exhibits a peculiar behavior that is involved in the phase transition.
Note that, while in the case of the ordinary first-order transition, $P_E(E|\beta)=\sum_E P(E,Q \ | \ \beta)$ in the energy region between the two most dominant $E$ values decreases, eventually approaching the shape shown in Fig. \ref{fig:sch_fo_and_so}(e), as $n$ increases.
On the other hand, in our model, $\big[P_E(E \ | \ \beta)\big]$ in the intermediate energy region remains unchanged with increased $n$.
In the vicinity of the special temperature $\beta_{\mathrm{fp}}$, the probabilistic contribution spans a broad range between $0.1 < [E]/n < 0.8$, as shown in Fig. \ref{fig:can_max_fp}(b). (In the figure, the distribution function $[D(\epsilon \ | \ \beta)]=(1/\Delta \epsilon)P_E(E \ | \ \beta)$, where $\epsilon=E/n$ and $\Delta \epsilon=1/n$, is shown for the collapse of the profile.)
Moreover, we observe the Monte Carlo trajectory of the instantaneous energy $E/n$ with the local updating of the variables $s_i$ in Eq. (\ref{eq:spin_variable}), using the Metropolis rule with the fixed temperature $\beta=\beta_{\mathrm{fp}}$ and without replica exchange or a reweighting potential: the result is shown in Fig. \ref{fig:can_max_fp}(c).
In that figure, the value of the instantaneous $E/n$ moves around smoothly in the broad $E$ region. This behavior is in contrast to that of the ordinary first-order transition, in which the instantaneous $E/n$ localizes in the vicinity of two peaks and the fluctuation becomes smaller with the increase in system size.
When $D(\epsilon)$ has the two peaks and the intermediate valley between them, it yields a discontinuous jump and hysteresis, which are involved in the ordinary first-order phase transition.
On the other hand, our model does not have an intermediate valley, which ensures contributions from the broad $E$ region.
Therefore, the dynamical property around the transition point $\beta_{\mathrm{fp}}$ differs significantly from the ordinary first-order phase transition.
The contributions from the broad region do not decay significantly with increased $n$.
Furthermore, this transition also differs from the so-called weak first-order transition, in the sense that the broad region spans a broad range of the $E$ region. In Fig. \ref{fig:can_max_fp}(b), the broad region even seems to expand as $n$ increases, whereas, in the case of the weak first-order transition, the weak valley spans a narrow $E$ region.
The disappearance of the curvature of $S(E)$ associated with the jump in the internal energy is partially similar to that of second-order transitions with the critical exponent $\alpha>0$.
However, note that, for the present transition, the region with $\partial^2 s/\partial \epsilon^2=0$ seems to have a finite range; however, this region remains only a single point in ordinary second-order transitions.
Therefore, in our case, the states are sampled from an energy region that is significantly broader than that of an ordinary second-order transition.
From these observations, we discard the possibility of the standard second-order phase transition.
We again emphasize that discontinuity in the macroscopic quantity is observed, as in the ordinary first-order phase transition; however, the dynamical properties involved in the phase transition, such as the discontinuous jump and hysteresis, differ significantly from those of the ordinary first-order phase transition.

The origin of the linear slope can be explained as follows.
First, the value of $\mathrm{mod}(N,d)$ can be up to $d-1$.
In fact, among all $d$ values in $2^{k}+1 \le d \le 2^{k+1}$, the resulting $\mathrm{mod}(N,d)$ typically lies in $2^{k} \le \mod(N,d) \le 2^{k+1}-1$.
Such $d$ has $E(d)=k$, according to the Hamiltonian (\ref{eq:max_model}).
Hence, $d$ in $2^{k}+1 \le d \le 2^{k+1}$ with $E(d)<k$ is (certainly but) rarely present.
For the same reason, in other regions with $k'$ ($>k$), the state $d$ rarely has $E(d)=k<k'$.
Thus, also among all possible states with $E=k$, the $d$ that lies in $2^{k}+1 \le d \le 2^{k+1}$ is typical.
This means that the total number of states with $E(d)=k$ is almost equal to $2^{k}$.
Since the resulting entropy with these states is $S(k)=k\log 2$, the number of states is roughly proportional to the value of $E$ itself.

\begin{figure}[h]
\begin{center}
\includegraphics[width=\figwidth]{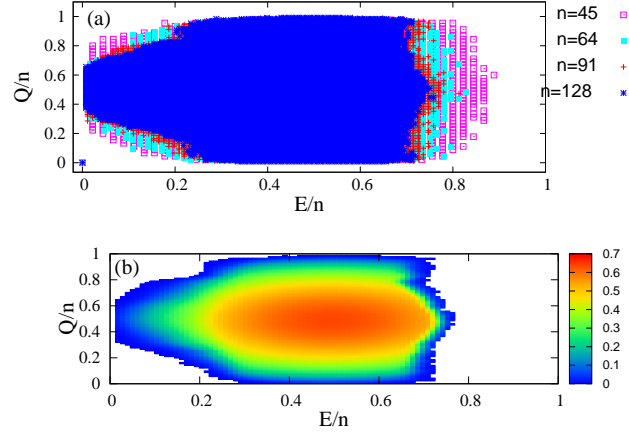} 
\caption{\label{fig:bhv_sum_model_3d}Distribution of $\log[W]/n$ on $(E/n,Q/n)$-plane for summation-based model. (a) Overplot of cases with $n = 45,64,91,$ and $128$. In the blank region, $W(E,Q)=0$ for all instances in the average. (b) $\log[ W(E,Q) ]/n$ for $n=91$. The value is represented using a color scale.}
\end{center}
\end{figure}

\begin{figure}[h]
\begin{center}
\includegraphics[width=\figwidth]{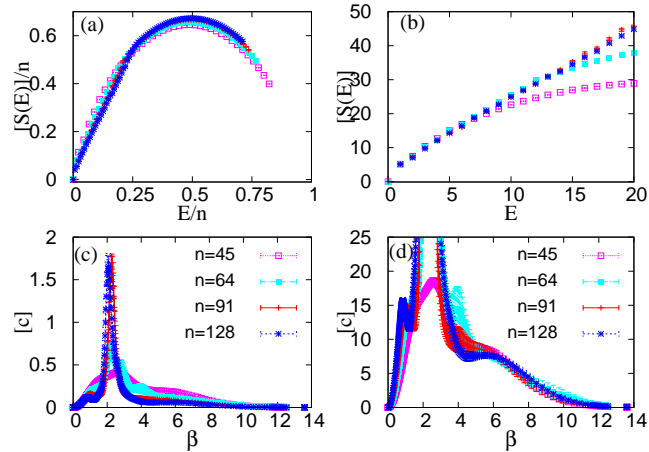} 
\caption{\label{fig:bhv_sum_model}$[S(E)]$ and $[c(\beta)]$ for summation-based model with $N=45, 64, 91, \mathrm{and} \ 128$. (a) $[S(E)]/n$ with respect to $E/n$, (b) low-energy region of $[S(E)]$ with respect to $E$, (c) $\beta$ dependence of $[c(\beta)]$, and (d) $\beta$ dependence of heat capacity $[C(\beta)]=n[c(\beta)]$.
}
\end{center}
\end{figure}

In order to confirm the generality of the behavior generated by our model, we investigate another model that is also constructed on the basis of the remainder and that has extensiveness, namely, the summation-based model [see Eq. (\ref{eq:sum_model})].
This model is the reduced version of the model proposed by Burges \cite{Bur} and experimentally implemented by Xu \cite{Xu} et al., with a liquid-crystal nuclear magnetic resonance (NMR) quantum processor.
The $[S(E)]$ of this model is shown in Figs. \ref{fig:bhv_sum_model}(a) and (b).
In a rather low-energy region, both the slope and the convex kink at $E=1$ are observed, which have identical properties to model (\ref{eq:max_model}), as seen in Figs. \ref{fig:ham_tmp_pfor}(a) and (c), respectively.
Furthermore, the results for $[c(\beta)]$ shown in Figs. \ref{fig:bhv_sum_model}(c) and (d) exhibit a sharp peak and a shoulder, corresponding to the first peak at $\beta_{\mathrm{fp}}$ and the smeared dip 
characterized by the convex kink in Fig. \ref{fig:sp_ht_max}, respectively.
Although common features that are expected to induce the same computational properties exist for both models, features that are not apparent in model (\ref{eq:max_model}) are also obtained.
In the high-$E$ region, the slope is connected to the clearly convex profile with a positive curvature.
Another convex kink in Fig. \ref{fig:bhv_sum_model}(a) also appears.
Correspondingly, for $[c(\beta)]$, another shoulder that decays with $O(n^{-1})$ is obtained.

\section{Summary and Discussion}
In this study, we have proposed two statistical mechanical models of the integer factorization problem and numerically investigated the static properties of these models.
The ground state of each model was isolated from other low-energy states by the Hamming distance $\simeq 0.5n$.
In the microcanonical entropy profile, we observed a peculiar shape with a convex kink and a linear slope.
These features were intensively discussed and related to a mechanism for marked changes in the sampling.
In particular, the linear slope of the microcanonical entropy exhibits a first-order phase transition in the macroscopic quantity, but a different behavior is observed in the energy distribution and the dynamics at the transition point.
A detailed interpretation of the behavior of the observed features as regards the computational hardness is left as a topic for further study.
However, at least, we can state that the linear shape has both aspects involved in the ordinary first- and second-order phase transitions in some sense, i.e., a discontinuous jump and disappearance of the microcanonical entropy curvature, respectively.
As observed from the modeled trajectory, the sampled state can easily move in the intermediate-energy region at the transition point.
This dynamical property differs significantly from the behavior of the ordinary first-order phase transition.
Concerning the phase transition, we conclude that the relaxation to the equilibrium state is significantly  easier than in the case with the ordinary first-order phase transition.

In the models proposed here, we observe only the behavior that confirms the hardness 
of the problem, at least in the context of the classical search problem.  However,
further investigation, or modification, of the model with regard to the unique classical computational property of the integer factorization problem, which differs from that of NP-complete problems, remains of considerable interest.
For example, we may find some trace of reduced computational hardness, to some extent, in the detailed observation of the structure of the free-energy landscape.
While the solution space of the 3-SAT problem has a hierarchically clustered feature, that of the number partitioning problem is merely random and no significant structure is observed.
It has been proposed that these differences may reflect some difference in manner in which the computational hardness is enclosed in these problems \cite{HMR}.

In addition, our present conclusion includes an interesting part as a classical counterpoint to quantum annealing.
In quantum annealing, the quantum phase transition plays an important role as regards the determination of the hardness of a computational problem. In the case of the first-order transition, the gap between the ground state and the lowest excited state vanishes exponentially with increasing system size, whereas, in the case of the second-order transition, the gap vanishes polynomially with increasing system size.
Indeed, the quantum first-order transition is observed in quantum Hamiltonians of NP-complete problems.
On the basis of this fact, the transition examined in this paper may represent the computationally remarkable feature of the prime factorization problem, namely, the fact that it can be solved in quasi-exponential time with effective classical algorithms.
To further explore this interesting subject, The behavior of the minimal gap of the Hamiltonian used in this paper is worth investigating with respect to the system size in the quantum annealing case \cite{Ozk,OKN,ONne1,ONne2}, for example, in the transverse field.
The energy levels of a classical (problem) Hamiltonian are reflected to the diagonal part of a quantum Hamiltonian in the case of quantum annealing.
Although the fictitious fluctuation differs from that of a thermal (probabilistic) case, the level arrangement may induce phenomena related to the latter.
In addition, in the case of the spin model, a certain similarity exists between the quantum phase diagram, with respect to the intensity of the transverse field, and the classical cases, with respect to the temperature \cite{NN}.
Therefore, in consideration of the classical phase transition,
the quasi-exponential dependence of the gap on the system size may be expected in quantum annealing, in the sense that the quasi-exponential dependence lies in the intermediate position between the polynomial and true exponential ones, although there is no reason to expect a polynomial one.

Finally, note that it is obviously one of the natural directions for future research to sophisticate the quantum Hamiltonian toward the faithful quantum-annealing version of Shor's algorithm, in which, ideally, the gap vanishes with polynomial dependence on the system size.

\begin{acknowledgment}
This work was supported by Grants-in-Aid Nos. 26610111 and 15H03699 from JSPS.
\end{acknowledgment}

\appendix
\section{Quenched and annealed average}
\begin{figure}[h]
\begin{center}
\includegraphics[width=\figwidth]{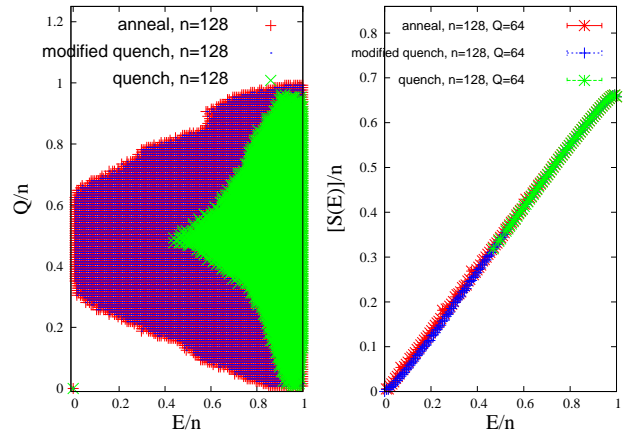}
\caption{\label{fig:q_a_mq}$\log[W(E,Q)]$ (anneal), $[\log(1+W(E,Q))]$ (modified quench), and $[S(E,Q)]$ (quench) with $n=128$, averaged over various instances of $N$. (a) The blank region represents the region with $\log[W(E,Q)]=-\infty$, $[\log(1+W(E,Q))]=0$, and $[S(E,Q)]=-\infty$. (b) Profile of each quantity with fixed $Q$: $Q=64$. The region with no points, particularly significant in $[S(E,Q)]$, means $[S(E,Q)]=-\infty$.
}
\end{center}
\end{figure}

The annealed average $\log[ \ W(E,Q) \ ]/n$ is taken to avoid the following problem caused by the quenched average with $W(E,Q)=0$.
In the case of the finite-size quenched average, $[S(E,Q)]=[\log W(E,Q)]$, even $W(E,Q)=0$ in a single instance immediately causes $[S(E,Q)]=-\infty$.
This problem is displayed in Fig. \ref{fig:q_a_mq}(a). In the $(E,Q)$ region where $[S(E,Q)]$ (quenched average) is not plotted, the instance with $W(E,Q)=0$ is included in the numerical sampling, namely, $[S(E,Q)]=-\infty$.
When $[W(E,Q)]$ is sufficiently large, it is considered that there is a bridge, although narrow, which allows the instantaneous state to move to other regions.
Even if some rare (or really single) instances with $W(E,Q) = 0$ are included, it is adequate to think that, in the average case over various instances, the instantaneous state is almost able to move passing through such a region.
Therefore, we adopted the annealed average $\log [W]$ for the discussion in the first part of Sect. \ref{sec:Numerical_Results}.

Furthermore, except for the region with a small $\log[W(E,Q)]$, where some instances with $W=0$ appear, the annealed average is a helpful quantity.
The comparison of quenched, annealed, and modified-quenched averages of $S(E,Q)$ with $n=91$ is shown in Fig. \ref{fig:q_a_mq}(b).
In the figure, the quantitative differences between $\log[W(E,Q)]$ and $[S(E,Q)]$ are small.

In addition, we note that, in the case of $[ S(E) ]=\big[ \log\big( \ \sum_{Q}\exp(S(E,Q)) \ \big) \big]$, which is focused after the latter part of Sect. \ref{sec:Numerical_Results}, $W(E)=0$, namely, $S(E)=-\infty$, in a certain $E$ region is not found, at least within the numerical sampling of instances.
Therefore, only the quenched average is shown for $S(E)$.

\section{Numerical simulation methods}
The numerical simulations performed for this study are based on the replica-exchange Monte Carlo method and the multiple histogram reweighting technique.
The methods used to obtain the results shown in Figs. \ref{fig:epsart}, \ref{fig:xygap_max}-\ref{fig:can_max_fp}, and \ref{fig:can_max_fp}(c) differ slightly from each other.

First, for Fig. \ref{fig:epsart}, each variable $s_i$ is updated using the Metropolis rule. That is, the accepting probability for local updating,  $p_{lu}$, is given by
\begin{eqnarray}\label{eq:p_lu}
p_{lu}=\mathrm{min}\{1,\exp\big(-\beta (H'-H)\big)\},
\end{eqnarray}
where $\beta$ is the physical inverse temperature and $H$ and $H'$ are the values of the Hamiltonian (\ref{eq:max_model}) for the configuration before and after the update, respectively.
The exchange of replicas with the indices $l$ and $l+1$ is performed using the Metropolis rule with the exchange probability $p_{exch}$, which is given as
\begin{eqnarray}\label{eq:p_exch}
p_{exch}=\mathrm{min}\{1,\exp\big((\beta^{(l)}-\beta^{(l+1)})(H^{(l+1)}-H^{(l)})\big)\}.
\end{eqnarray}
For Fig. \ref{fig:epsart}, to investigate the efficiency for a local search problem, the Hamming distance to the solution, $Q$, is not taken into account for $p_{lu}$ and $p_{exch}$ in this situation.
The first passage time is given as the number of time steps required until the solution is found by any of the replicas.

Next, for Fig. \ref{fig:xygap_max}-\ref{fig:can_max_fp}, $Q$ is introduced in order to accelerate the sampling, since the ground state and other low but positive energy states are largely separated.
To reduce the difficulty in sampling with the original Hamiltonian, the new cost function $V(E,Q)$ is introduced to the replica exchange Monte-Carlo method.
Acceptance probabilities for local updating and replica exchange with $V(E,Q)$, $p_{lu}^{(V)}$, and $p_{exch}^{(V)}$ are given respectively in a similar manner as Eqs.(\ref{eq:p_lu}) and (\ref{eq:p_exch}), namely,
\begin{eqnarray}
p_{lu}^{(V)}=\mathrm{min}\{1,\exp\big(-\gamma (V'-V)\big)\}, \\
p_{exch}^{(V)}=\mathrm{min}\{1,\exp\big((\gamma^{(l)}-\gamma^{(l+1)})(V^{(l+1)}-V^{(l)})\big)\}.
\end{eqnarray}
The form of $V(E,Q)$ is prepared to reproduce the approximated value of the microcanonical entropy $\tilde{S}(E,Q)$, namely, $V(E,Q)=\tilde{S}(E,Q)$.
Note that $E$ itself is still given by the value of the Hamiltonian (\ref{eq:max_model}) or (\ref{eq:sum_model}).
Only the acceptance ratio, namely, the weight in the stational distribution in the simulation, is modified.
The method proposed here is similar to the method in the papers of Mitsutake et al., \cite{MSO,MO} in the sense that it is a hybrid of the multicanonical \cite{BN,WL} or entropic sampling \cite{Lee} and the replica exchange.
However, the proposed method is slightly different from them.
In the method of this paper, each simulation weighted by a different factor $\exp(-\gamma^{(l)} S(E,Q))$ samples $(E, Q)$-states inhabiting different value regions of the microcanonical entropy itself.
At $\gamma >1$, entropically rare states are preferably sampled, while typical states are preferred at $\gamma \le 1$.

In the case of Hamiltonian (\ref{eq:max_model}) or (\ref{eq:sum_model}), when the MC simulation is performed with the weight $\exp(-\beta E)$, there is a large barriar between the ground state and other states with $Q=1$.
However, with the weight $\exp(-\gamma^{(l)} S(E,Q))$, the barrier is considerably reduced.

\begin{figure}[h]
\begin{center}
\includegraphics[width=\figwidth]
{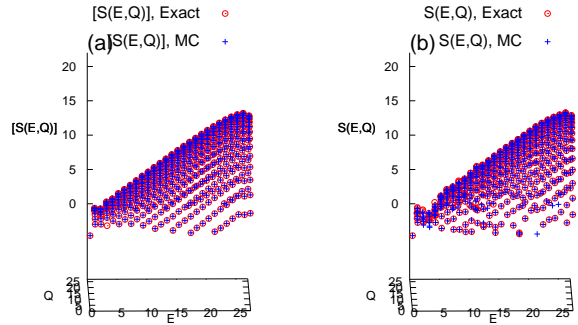}
  \caption{\label{fig:exact_hikaku}Comparison of $S(E,Q)$ with $n=27$ computed by Monte Carlo simulation (blue) and exact enumeration (red). (a) $[S(E,Q)]$ averaged over instances of $N$. (b) $S(E,Q)$ with $N=35176793\times229429217$.
}
  \label{fig:hik_1}
\end{center}
\end{figure}

\begin{figure}[h]
\begin{center}
\includegraphics[width=\figwidth]
{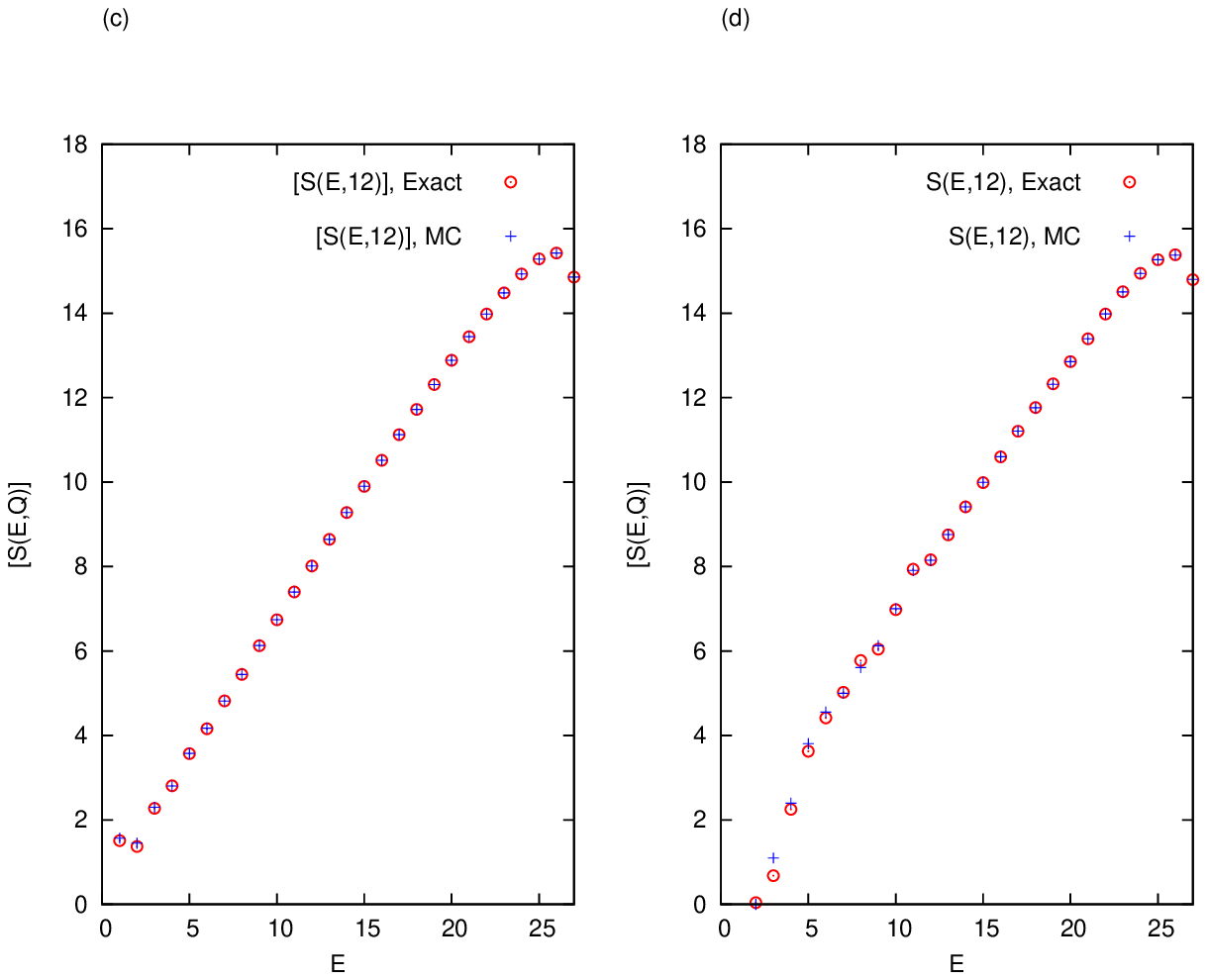}
  \caption{\label{fig:hik_2}Comparison of $S(E,Q)$ with $n=27$ and $Q=12$ computed by Monte Carlo simulation (blue) and exact enumeration (red). (a) $[S(E,12)]$ averaged over instances of $N$. (b) $S(E,12)$ with $N=35176793\times229429217$.
}
\end{center}
\end{figure}

\begin{figure}[h]
\begin{center}
\includegraphics[width=\figwidth]
{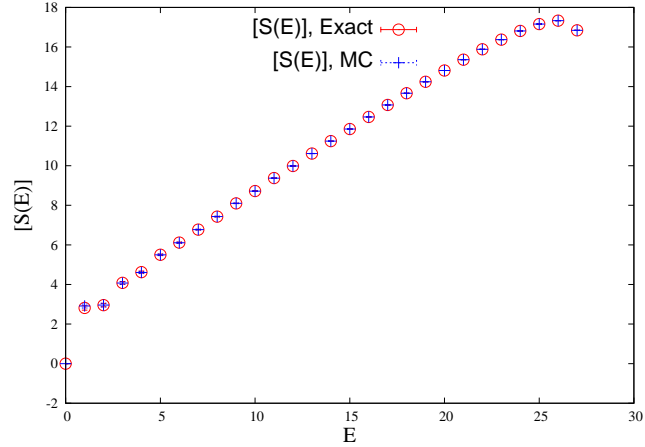}
  \caption{\label{fig:hik_3}Comparison of $[S(E)]=\big[\log\big(\sum_Q W(E,Q)\big)\big]$ with $n=27$ computed by Monte Carlo simulation (blue) and exact enumeration (red).
}
\end{center}
\end{figure}

To examine the validity of this method, the comparison of the results with the Monte Carlo simulation to those with exact enumeration is shown in Figs. \ref{fig:hik_1}-\ref{fig:hik_3}.
As shown in Fig. \ref{fig:hik_1}(b) and \ref{fig:hik_2}(b), small differences are still found for the small-$S(E,Q)$ region with some instances.
However, under most instances and in a large part of the $(E,Q)$-plane, the two results are in good agreement.
Hence, at least for a small $n$, the Monte Carlo results for $\log [W(E,Q)]$, $[S(E,Q)]$, or $[S(E)]$ are in good agreement with the exact ones in almost all regions, as shown in Figs. \ref{fig:hik_1}(a), \ref{fig:hik_2}(a), and \ref{fig:hik_3}.

\begin{figure}[h]
\begin{center}
\includegraphics[width=\figwidth]{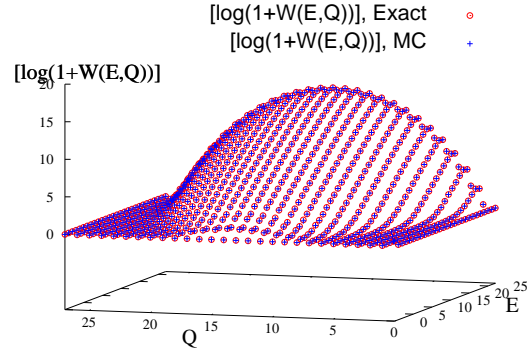}
\caption{\label{fig:potential}$[\log (1+W(E,Q))]$ with small size, $n=27$, computed by exact enumeration and Monte Carlo simulation.
}
\end{center}
\end{figure}

In this method, the cost function $V(E,Q)$ is prepared as follows.
Starting from a small size $n_1$, we first compute $[\log(1+W(E,Q))]$ by the exact computation as shown in Fig. \ref{fig:potential}.
Here, we note $[\log(1+W(E,Q))]$ and $V(E,Q)$ with $n_1$ as $[\log(1+W(E,Q; n_1))]$ and $V(E,Q; n_1)$, respectively.
Then, we use the homothetic of $[\log(1+W(E,Q; n_1))]$ as the potential for a larger system size $n_2$, $V(E,Q ; n_2)$, by substituting it as $V(E,Q; n_2) = [\log (1+W(rE,rQ; n_1))]$, where $r=n_2/n_1$, to compute $[\log(1+W(E,Q; n_2))]$ and $[S(E,Q ; n_2)]$ by the Monte Carlo method.
Thereafter, we recursively perform this procedure and gradually increase the system size of the simulation with $V(E,Q; n)$ in the same manner.

The histograms sampled with each replica, $h_l(E,Q)$, are integrated to estimate the number of states $W(E,Q)$ by the multiple histogram reweighting method.
$W(E,Q)$ is obtained as
\begin{eqnarray}\label{eq:recursion}
W(E,Q)=\frac{\sum_{l}h_l(E,Q)}{\sum_{l}\big(\frac{\omega_l(E,Q)}{z_l}\sum_{E=0}^{n}\sum_{Q=0}^{n}h_{l}(E,Q)\big)},
\end{eqnarray}
where
\begin{eqnarray}
\omega_l(E,Q)&=&\exp\big(-\gamma_l V(E,Q)\big), \label{eq:omega}\\
z_l&=&\sum_{E=0}^{n}\sum_{Q=0}^{n}W(E,Q)\omega_l(E,Q).
\end{eqnarray}
Equation (\ref{eq:recursion}) has a recursive form because $z_l$ on the right-hand side, explicitly written as (\ref{eq:omega}), includes $W(E,Q)$.
Equations (\ref{eq:recursion}) and (\ref{eq:omega}) are interpreted as the most-likelihood inference and equivalent to the minimization of the log-likelihood function
\begin{eqnarray}
L(W(E,Q))=&&\sum_{l}\sum_{E=0}^{n}\sum_{Q=0}^{n}\Big\{ h_{l}(E,Q)\Big( \log W(E,Q)+\log \omega_l(E,Q)-\log z_l\Big) \ \Big\}.
\end{eqnarray}
Equation (\ref{eq:recursion}) is originally derived in the study of Ferrenberg and Swendsen \cite{FS}, and the inference interpretation is proposed in another study \cite{BK}.

Finally, as for Fig. \ref{fig:can_max_fp}(c), we use the same local updating as Eq. (\ref{eq:p_lu}) with $\beta=\beta_{\mathrm{fp}}$ without replica exchange or any reweighting potential.

\end{document}